\documentstyle[pra,aps,epsfig,floats]{revtex}

\begin{document}
\draft
\twocolumn[\hsize\textwidth\columnwidth\hsize
\csname@twocolumnfalse\endcsname

\title{Multi-mode Interferometer for Guided Matter Waves}
\author{Erika Andersson$^1$,
Tommaso Calarco$^{2}$, Ron Folman$^3$,
Mauritz Andersson$^4$, Bj\"orn Hessmo$^5$, J\"org Schmiedmayer$^3$
}

\address{
 $^1$ Department of Physics, Royal Institute of Technology,
 SE-10044 Stockholm, Sweden \\
 (Present address: Department of Physics and Applied Physics,
University of Strathclyde, Glasgow G4 0NG, Scotland)\\
 $^2$ Institut f\"ur Theoretische Physik, Universit\"at Innsbruck,
 A-6020 Innsbruck, Austria\\
 European Centre for Theoretical Studies in Nuclear Physics and Related
 Areas, 38050 Villazzano (TN) Italy\\
 $^3$ Physikalisches Institut, Universit\"at Heidelberg, D-69120
 Heidelberg Germany\\
 $^4$Department of Quantum Chemistry, Uppsala University, S-75120
 Uppsala, Sweden\\
 $^5$ Dept. of Microelectronics and Information Technology,
Royal Institute of Technology, S-164 40 Kista, Sweden}

\date{\today}

\maketitle

\begin{abstract}
Atoms can be trapped and guided with electromagnetic fields,
using nano-fabricated structures. We describe the fundamental
features of an interferometer for guided matter waves, built of
two combined Y-shaped beam splitters. We find that such a device
is expected to exhibit high contrast fringes even in a multi-mode
regime, analogous to a white light interferometer.
\end{abstract}


\vskip1pc ] \narrowtext Interferometers are very sensitive
devices, and have provided both insights into fundamental
questions and valuable instruments for applications. The
sensitivity of matter-wave interferometers \cite{MWI} has been
shown to be much better than that of light interferometers in
several areas such as the observation of inertial effects
\cite{zero}. Because of this high sensitivity, interferometers
have to be built in a robust manner to be applicable. This could
be achieved by guiding the matter waves with microfabricated
structures, and by integration of the components into a single
compact device, as has been done with optical devices.

In this Letter, we describe such an interferometer for matter
waves propagating in a time-independent guiding potential, in
analogy with propagation of light in optical fibers (for recent
microtrap proposals see \cite{history}). This interferometer has
the surprising feature that interference is observed even if many
levels in the guide are occupied, as is the case when the source
is thermal, or for cold fermions, even below the Fermi
temperature. The multi-mode interferometer is built by combining
two Y-shaped beam splitters \cite{cass}, capable of coherently
splitting or recombining many incoming transverse modes, and
arranging the interferometer geometry so that all the different
transverse states give the same phase shift pattern. The beam
splitters are formed by splitting the atom guiding potential
symmetrically into two identical output guides. This was recently
demonstrated on an atom chip \cite{cass,four}; alternatively,
this may be realized by optical confinement \cite{french}.

The analysis is done in two dimensions \cite{2d}, similar to
solid state electron interferometers \cite{electron}.
The shape of the guiding potential in the transverse direction $x$ changes
with the longitudinal coordinate $z$, from a single harmonic well of
frequency $\omega$ to two wells of frequency $2\omega$ separated by a
distance $d(z)$, and then back again to a single guide (see Fig. 1a and b).
\begin{figure}[tbh]
    \vspace{-0.5truecm}
    \begin{center}\hspace{0mm}\mbox{\input epsf
\epsfxsize\columnwidth\epsfbox{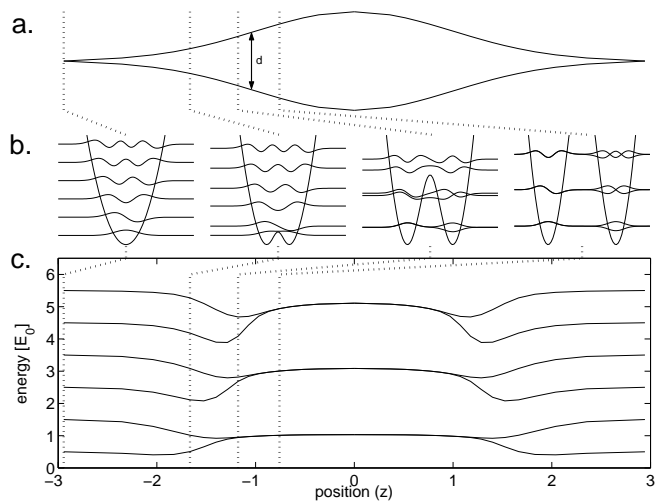}}\end{center}
    \caption{
    The guided matter wave interferometer.
    {\em a})  Two beam splitters are joined together to form the
    interferometer.
    {\em b}) Transverse eigen-functions of the guiding potentials in
    various places along the first beam splitter.
    When the two outgoing guides are separated far
    enough, so that no tunneling between left and right occurs,
    the symmetric and the antisymmetric states become pairwise
    degenerate.
    {\em c}) Energy eigenvalues for the lowest transverse
    modes as they evolve along the interferometer.}
    \label{fig:schema}
\end{figure}
Let us consider a particle entering in the transverse ground
state and with longitudinal kinetic energy $E_{\rm kin}$. In the
limit where the atomic transverse motion in the guide (related to
$\omega$ and the ground state size) is very fast with respect to
the rate of transverse guide displacement $\frac{d}{dt}d(z)$ as
seen by the moving atom (related to the beam splitter opening
angle $\frac{d}{dz}d(z)$ and $E_{\rm kin}$), the particle will
adiabatically follow the lowest energy level throughout the
interferometer. This corresponds to the condition $E_{\rm
kin}[\frac{d}{dz}d(z)]^2\ll\hbar\omega$. As it can be seen in Fig.
1b, when the two guide arms are far apart (i.e., when
$d(z)\gg\sqrt{\hbar/m\omega}$), any eigenstate of the incoming
guide potential evolves into a superposition of eigenstates
$|n\rangle_l$ and $|n\rangle_r$, corresponding to the left and
right arms, respectively:
\begin{eqnarray}
\label{adiabfollow}
    &|2n\rangle\longrightarrow\frac 1{\sqrt
    2}\left[|n\rangle_l+|n\rangle_r\right] \nonumber \\
    &|2n+1\rangle\longrightarrow\frac 1{\sqrt
    2}\left[|n\rangle_l-|n\rangle_r\right].
\end{eqnarray}
Levels $|2n\rangle$ and $|2n+1\rangle$ become practically
degenerate inside the interferometer when the arms are widely
split, as Fig. 1c shows. For example, it is easy to visualize how
a transverse odd state $|1\rangle$ splits in the middle to become
a superposition of two ground states having between them a $\pi$
phase, while the even state $|0\rangle$ does the same, but with a
$0$ relative phase. As the states become degenerate, an
asymmetric perturbation (e.g., a differential phase shift) will
couple the odd and even symmetries, thus inducing a mixing between
the two states (see also \cite{history}).
Numerical two-dimensional wave-packet calculations for the lowest
35 modes \cite{SplitOp} confirmed that no transitions between
transverse states occur, as long as the motion is adiabatic
in the sense discussed above, and as long as no asymmetric
perturbation is present. In an ideal case, coherent splitting and
recombination for all transverse modes can be achieved
\cite{tunnelingBS}.

As mentioned, mixing within pairs of degenerate states inside the
interferometer can occur if the wavefunction experiences a phase
difference between the two arms. Let us therefore introduce a
phase shift $\Delta \phi$ between the interferometer arms, either
by making one arm longer by $\Delta l$, giving $\Delta\phi
=k\Delta l$, where $\hbar k$ is the momentum in the longitudinal
($z$) direction, or by applying an additional potential $U$,
resulting in $\Delta\phi =\frac{m}{\hbar^2 k}\int U dz$. Both
phase shifts are independent of the transverse state in the
guide, and they are dispersive, meaning that the resulting phase
shift
depends on $k$. 
In the following, we shall refer to the phase shift caused by a
path length difference $\Delta l$.

\begin{figure}[tb]
    \vspace{-0.25truecm}
    \begin{center}\hspace{0mm}\mbox{\input epsf
\epsfxsize\columnwidth\epsfbox{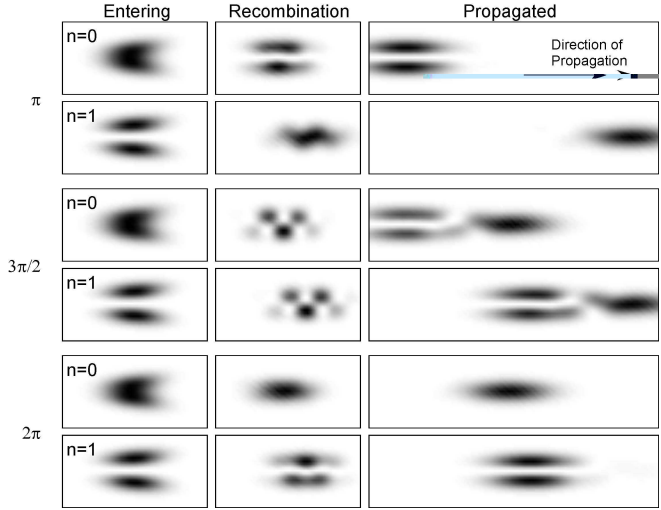}}\end{center}
    \caption{Probability distributions of wave packets in the
interferometer. The probability distributions are shown just
before entering, right after exiting the interferometer, and
after a rephasing time $t$. The incident wave packet is in the
transverse ground state ($n=0$) or first excited state ($n=1$),
the phase shift between the interferometer arms is $\pi$,
$\frac{3 \pi}{2}$ or $2\pi$. In this calculation, the wave packet
is relatively narrow in momentum space, so that the whole wave
packet obtains approximately the same $k$-dependent phase shift
$\Delta\phi$. If $\Delta\phi$ is $2\pi$, the wave packet exits in
the same transverse state as it entered. If it is $\pi$, $n=0$
exits as $n=1$ and vice versa. For $\Delta\phi =3\pi /2$, the
wave packet will exit in an equal superposition of the transverse
ground state and first excited state. The separation of these two
outgoing components after the rephasing time is due to energy
conservation. In this example, the wave packets are sufficiently
narrow in longitudinal momentum space, so that the whole wave
packet acquires approximately the same $k$-dependent phase shift.
The numerical calculation is done by solving the time-dependent
Schr\"odinger equation in two dimensions using the split-operator
method,
for realistic guiding potentials.}
    \label{fig:IFMsimple}
\end{figure}

Following equation \ref{adiabfollow} and its time inverse, an
incoming even state is transformed as follows while transversing
through the two beam splitters BS1 and BS2:
\begin{eqnarray}
|2n\rangle&\stackrel{BS1+\Delta\phi}\longrightarrow
&\frac{1}{\sqrt{2}}[e^{i\Delta\phi/2}|n\rangle_{l}+
e^{-i\Delta\phi/2}|n\rangle_{r}]\nonumber\\
&\stackrel{BS2}\longrightarrow &
\frac{1}{\sqrt{2}}[\cos(\Delta\phi/2)|2n\rangle+
i\sin(\Delta\phi/2)|2n+1\rangle].
\end{eqnarray}
By taking into account the analogous transformation rule for the nearby odd
state, we find that we can describe the interferometer
in terms of a matrix
\begin{equation}
   \begin{array}
   [c]{cc}%
   {1\over\sqrt{2}}\left(
   \begin{array}
   [c]{ll}%
   \cos(\Delta \phi/2) & i\sin(\Delta \phi/2)\\
   i\sin(\Delta \phi/2) & \cos(\Delta \phi/2)
   \end{array}
   \right)
   \end{array}
\label{eq:matrix}
\end{equation}
in the basis of the two incoming and outgoing transverse
eigenmodes $|2n\rangle$ and $|2n+1\rangle$. Figure
\ref{fig:IFMsimple} shows the above qualitative behavior in an
actual simulation.

Let us now discuss the longitudinal degree of freedom. Consider a
very cold wave packet with longitudinal kinetic energy
$E_{kin}=k^2 \hbar^2/2m$, where $k$ is the mean longitudinal
momentum of the wave packet. In the transverse direction it
occupies the transverse ground state $|0\rangle$ of the incoming
guide. We have chosen the ground state energy in the
interferometer arms, $\hbar\omega^\prime$, to be twice that of
the input and output guides, $\hbar\omega$ (see Fig.
\ref{fig:schema}). This implies that if the longitudinal kinetic
energy is too small, $E_{kin}<\frac{1}{2}\hbar(\omega^\prime
-\omega)=\frac{1}{2}\hbar\omega$, the wave packet will be
reflected already at the first beam splitter. For
$E_{kin}>\frac{1}{2} \hbar\omega$, it will split between the two
interferometer arms to occupy the transverse ground states in both
arms, slowing down in the longitudinal direction to ensure energy
conservation. If the wave packet experiences a phase shift
$\Delta\phi =k\Delta l$ 
in one of the arms, part of it will, after the recombination at
the second beam splitter, exit in the first excited transverse
state $|1\rangle$ of the outgoing guide, as described by matrix
(\ref{eq:matrix}), again provided the longitudinal kinetic energy
is large enough. The phase shift is dependent on $k$, and thus
the different $k$ components of the wave packet will obtain
different phase shifts. The components with phase shifts close to
$2N\pi$ will exit in the transverse ground state, and those with
phase shifts close to $(2N+1)\pi$ will exit in the first excited
state. If $\frac{1}{2}\hbar\omega < E_{kin}<\hbar \omega$, the
transition to the first excited transverse state will be
forbidden, and this part of the wave packet will be reflected. If
the kinetic energy of the wave packet fulfills $E_{kin}\gg \hbar
\omega$, transitions to the first excited transverse state at the
recombination beam splitter are possible, and no sizeable back
reflection occurs. The interference pattern may then be observed
by looking at the populations in the states $|0\rangle$ and
$|1\rangle$ in the outgoing guide, as presented in figure
\ref{fig:IFMsimple}.

We note that the part of the wave packet making the transition to
the state $|1\rangle$ will lose kinetic energy to compensate for
the additional transverse energy $\hbar \omega$ needed for the
transition. Therefore the part of the wave packet in state
$|1\rangle$ will travel slower than the part in $|0\rangle$, by
an amount $\Delta v\simeq\omega/k$.  As explicitly shown in the
third column of Fig. \ref{fig:IFMsimple} for $\Delta\phi =3\pi
/2$, the two outgoing components will separate longitudinally.
Similarly, a wave packet entering in the first excited transverse
state, acquiring an odd phase shift $(2N+1)\pi$, will exit in the
ground state, gaining potential energy, and propagating faster
than the components exiting in the first excited transverse
state. To observe this separation between the two outgoing parts
of the wave packet, one has to introduce a pulsed source, which
is what we will assume from now on.

The reasoning above is easily extended to higher transverse modes.
Components of a wave packet acquiring an even phase shift will
exit in the same transverse state as they entered, while
components acquiring an odd phase shift will exit in the
neighboring transverse state ($|2n \rangle \rightarrow |2n+1
\rangle$,  or $|2n+1 \rangle \rightarrow |2n \rangle$), lose
(gain) kinetic energy accordingly, and propagate slower (faster)
by the {\em same} $\Delta v$, independent of the transverse
energy level. Thus, for a system of temperature $T$, filling
$2n+1$ transverse modes, the interferometer is actually composed
of $n$ disjunct interferometers, all giving rise to the same
longitudinal pattern.

\begin{figure}[tb]
    \vspace{-0.25truecm}
    \begin{center}\hspace{0mm}\mbox{\input epsf
\epsfxsize\columnwidth\epsfbox{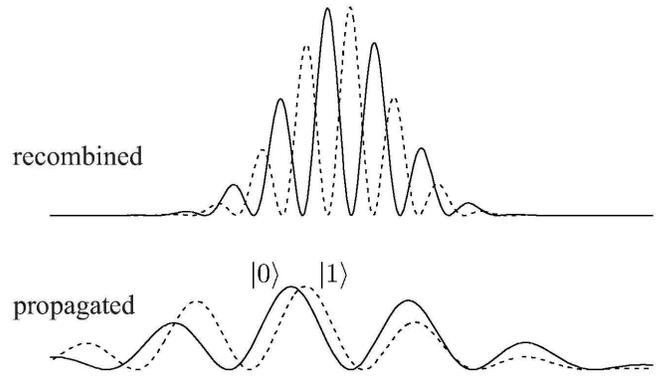}}\end{center}
    \caption{
    The rephasing of the interference patterns created
    by a wave packet entering in the ground state $|0\rangle$.
    The full line shows the probability distribution of the transverse
    ground state $|0\rangle$, and the dashed line shows that of the
    first excited state $|1\rangle$.
    The wave packet width is chosen to have a phase difference
    greater than $2 \pi$ between its front end and tail.
    The top graph shows the wave packet
    just after recombination at the end of the interferometer,
    the lower graph after propagation, when the $|0\rangle$ part
    has caught up with the $|1\rangle$ part. The wave packet has
    also spread out.}
    \label{fig:IFMrephasing}
\end{figure}

If the energy spread of the longitudinal wave packet is large
enough, $\Delta k\gg\pi/\Delta l$, a longitudinal interference
pattern will form within the wave packet, as shown in
Fig.~\ref{fig:IFMrephasing}. This pattern, shown here for the
states $|0\rangle$ and $|1\rangle$, will be the same for all
transverse states $|2n\rangle$ and $|2n+1\rangle$. The two
density patterns will, at the exit of the interferometer, add up
to the same wave packet shape one would expect if the
interferometer was not there \cite{IFMeff}. Looking at the total
wave packet, not distinguishing between the different transverse
levels, one only sees its envelope. The two transverse state
components will, however, propagate with different velocities, as
outlined above, and after some time they will rephase as shown in
Fig.~\ref{fig:IFMrephasing}. This will happen once enough time
has passed for $|\Delta v|$ to overcome the difference in the
position of the pattern peaks. Since $|\Delta v|$ is independent
of the incoming transverse mode, all patterns will re-phase at
the same time and position, and a multi transverse mode operation
could be achieved.

In an actual multi mode experiment, the simplest input state will
be a thermal atomic cloud. Such a state is described by a mixed
density matrix.
The longitudinal states are obtained in the following way: At the
start of the experiment, we imagine a thermal atomic cloud
trapped in both transverse and longitudinal directions, with
transverse confinement by the same trapping frequency $\omega$ as
in the incoming guide. In the longitudinal direction, the initial
trap is approximated by an infinitely high box of width $L$. The
states we consider are the eigenfunctions of this trap. At time
$t=0$, one wall of the box is opened up and the longitudinal
states start propagating in the positive $z$ direction according
to their momentum distributions. Each longitudinal state
 can be approximated by a wave packet with mean
momentum $k=(n+1)\pi/L$ and momentum spread $\Delta
k=2\sqrt\pi/L$ \cite{wp}.

\begin{figure}[tb]
\vspace{-0.5truecm}
    \begin{center}\hspace{0mm}\mbox{\input epsf
\epsfxsize\columnwidth\epsfbox{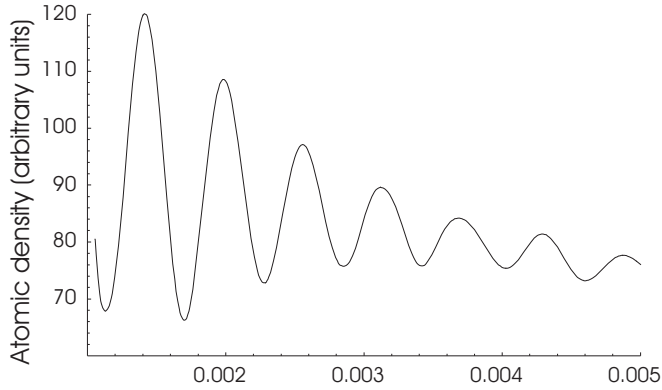}}\end{center}
    \vspace{-0.5truecm}
    \caption{
    The longitudinal interference pattern (Li atoms) as a function of $z$.
    The interferometer is $1$mm long with the source at its beginning.
    The transverse trapping frequency is $\omega=0.1\times 10^6$s$^{-1}$ and the length
    of the source trap is $L=100\mu$m. The path difference between
    the arms is $\Delta l=2\mu$m. The time
    elapsed since the release of the atoms from the trap is $20$ms. The
    periodicity of the interference pattern is about $550\mu$m, consistent with
    $\cos^2 [m \Delta l z/(2 \hbar t)]$, the analytically calculated modulation for a
    single mode in the limit
    of an initially narrow width source wave packet and a sufficiently long propagation
    time.
    The intensity is linearly dependent on the intensity of the source.}
    \label{pattern}
\end{figure}

We carry out a numerical calculation of the interference pattern
after the interferometer, starting from a thermal state as
described. We take into account the $k$-dependent phase shifts
acquired by the plane wave components of the longitudinal wave
packets, resulting in transitions between different transverse
states according to Eq. (\ref{eq:matrix}). The plane wave
components are slowed down or sped up, if necessary, to ensure
energy conservation. The emergence of an interference pattern,
despite the incoherent sum over the transverse and longitudinal
states, is confirmed by this numerical calculation. Figure
\ref{pattern} gives an example of an interference pattern
obtained with a typical ``hot'' atom ensemble with a temperature
of $200$ $\mu$K and guides with a trap frequency of
$10^5$s$^{-1}$, already realized on an atom chip \cite{ron,AtCh}.
Even though hundreds of transverse levels are populated, high
contrast fringes are observed.


To summarize, we have shown that a multi-mode interferometer may
be realized in a two-dimensional geometry, using two symmetric
Y-shaped beam splitters. Furthermore, though beyond the scope
of the present work, we expect three-dimensional evolution to
exhibit qualitatively the same behavior under certain conditions.
As Y-shaped microfabricated beam splitters have been realized, we
expect the road to be open for the experimental realization of
robust guided matter wave interferometers.

We would like to thank Peter Zoller for enlightening discussions.
R.F. is grateful to Yoseph Imry for his insight into mesoscopic
systems. E.A. would like to thank Helmut Ritsch for his kind
hospitality. This work was supported by the Austrian Science
Foundation (FWF), project SFB 15-07, by the Istituto Trentino di
Cultura (ITC) and by the European Union, contract Nr.
IST-1999-11055 (ACQUIRE), HPMF-CT-1999-00211 and
HPMF-CT-1999-00235.


\begin{references}
\bibitem{IFMeff}
    This is only approximately true, because inside the interferometer
    there are different propagation velocities due to the different
    potential energies of even and odd incoming states.  In the
    calculations, this small effect is included.
\bibitem{wp}
In this model, and for a reasonable source size $L$, the wave
packet energy spread will be relatively small and no complete
pattern of several oscillations is expected to be observable
within a single packet. This however, does not alter the final
result as no coherence between different longitudinal
$k$-components is needed. The only coherence evoked in the
interferometer is that between the right and left paths, which in
turn requires that a single wave packet have a coherence length
longer than the path length difference $\Delta l$. The number of
observed oscillations will then only depend on the longitudinal
energy spread of the source.
\end{references}
\end{document}